\documentclass[12pt,epsf]{article}

\usepackage[dvips]{graphicx}

\newcommand{\beq}{\begin{equation}}
\newcommand{\eeq}{\end{equation}}
\newcommand{\bea}{\begin{eqnarray}}
\newcommand{\eea}{\end{eqnarray}}

\begin{document}
\thispagestyle{empty}
\vspace*{-15mm}
\baselineskip 1pt
\begin{flushright}
\begin{tabular}{l}
{\bf OCHA-PP-277}\\
{\bf October 2008}\\
\end{tabular}
\end{flushright}
\baselineskip 24pt
\vglue 10mm

\begin{center}
{\LARGE\bf
Factorization of number into prime numbers viewed as decay of particle into elementary particles conserving energy 
}
\vspace{7mm}

\baselineskip 18pt
{\bf Akio SUGAMOTO}
\vspace{2mm}

{\it
   Department of Physics, Ochanomizu University, Tokyo 112-8610, Japan}\\
\vspace{10mm}
\end{center}
\begin{center}
{\bf Abstract}\\[7mm]
\begin{minipage}{14cm}
\baselineskip 16pt
\noindent
Number theory is considered, by proposing quantum mechanical models and string-like models at zero and finite temperatures, where the factorization of number into prime numbers is viewed as the decay of particle into elementary particles conserving energy. 
In these models, energy of a particle labeled by an integer $n$ is assumed or derived to being proportional to $\ln n$.  

The one-loop vacuum amplitudes, the free energies and the partition functions at finite temperature of the string-like models are estimated and compared with the zeta functions.  The $SL(2, {\bf Z})$ modular symmetry, being manifest in the free energies is broken down to the additive symmetry of integers, ${\bf Z}_{+}$, after interactions are turned on.
In the dynamical model existing behind the zeta function, prepared are the fields labeled by prime numbers. 

On the other hand the fields in our models are labeled, not by prime numbers but by integers.  Nevertheless, we can understand whether a number is prime or not prime by the decay rate, namely by the corresponding particle can decay or can not decay through interactions conserving energy.  Among the models proposed, the supersymmetric string-like model has the merit of that the zero point energies are cancelled and the energy levels may be stable against radiative corrections.

\end{minipage}
\end{center}

\newpage
\baselineskip 18pt
\def\thefootnote{\fnsymbol{footnote}}
\setcounter{footnote}{0}

\section{Introduction}
Recently, the long-standing unsolved conjectures of Fermat conjecture, Shimura-Taniyama conjecture and Sato-Tate conjecture are successively solved. (See \cite{Kato}, \cite{Taylor} and the references cited therein.)  The lecture by Kato \cite{Kato} and the reading of the original paper by Riemann  \cite{Riemann} are enough to move the author strongly towards considering the number theory, even though he is completely ignorant of the field.   

Motivation of this paper is as follows.  

In number theory, how a number factorizes into prime numbers is a key issue, while in particle physics how a particle decays into elementary particles is also a key issue.  These two key issues are intimately related, if we identify the energy $E_{n}$ of a particle labeled by a positive integer $n=1, 2, 3, \dots$ is proportional to $\ln n$.  Namely,
\beq
E_n=\omega \ln n,
\eeq
where $\omega$ is a unit of energy.

Then, factorization of a number into prime numbers \{$p_1, p_2, p_3, \dots$\},
\beq
n=(p_1)^{c_{1}}( p_2)^{c_{2}}(p_3)^{c_{3}}\cdots, \label{factorization into primes}
\eeq
can be viewed as the energy conservation law,
\beq
E_n=c_{1}E_{p_{1}}+c_{2}E_{p_{2}}+c_{3}E_{p_{3}}+\cdots, \label{energy conservation for primes}
\eeq
where $c_1, c_2, c_3, \dots$ are positive integers.

In this sense, a particle (labeled by) $n$ can be a bound state made up of $c_1$ particle $p_1$, $c_2$ particle $p_2$, $c_3$ particle $p_3$, and so on, and it can decay into elementary particles with smaller energies.  In this decay process the total energy should be conserved.  Therefore, it may be possible to understand number theory in terms of (particle) physics terminologies.  

Even though the factorization of number and the particle's decay following energy conservation law are nothing but the paraphrasing of each other, the physical intuitions on energy, free energy, decay width, unitarity and the optical theorem, supersymmetry, {\it etc.} may be useful in number theory.  Also the physical treatment on the symmetry and its breaking of the dynamical system may be helpful there. 

Purpose of this paper is to attempt for rewriting number theory in terms of the quantum mechanical theory and the two dimensional string-like quantum field theory, at zero as well as finite temperatures.

\section{Quantum model with infinitely many species of particles} 
In quantum theory, one species of particle, labeled by $n~(=1, 2, 3, \dots)$, is described by a field $\phi_{n}(t)$ depending on time $t$.  If the Lagrangian $L$ is given for a infinite set of fields \{$\phi_{n}(t)$\} (n=1,2,3,\dots), the dynamics between particles, or how (with what probablity) each particle decays into other particles is known.  The simplest quantum model ``Model $A$" is the non-relativistic bosonic theory with cubic coupling $\lambda$, whose Lagrangian is
\bea
	L_{A}&=&\sum_{n}\left(\overline{\phi}_n(t)i \frac{d}{dt}\phi_{n}(t)-E_n^b~\overline{\phi}_n(t)\phi_{n}(t)\right) \nonumber \\ 
	&-&\sum_{n_1, n_2, n_3}\lambda \left(\overline{\phi}_{n_1}(t)\phi_{n_2}(t)\phi_{n_3}(t)+\phi_{n_1}(t)\overline{\phi}_{n_2}(t)\overline{\phi}_{n_3}(t)\right).
\eea
Here, the ``overline" means the complex conjugation and the affix ``$b$" indicates ``bosonic".
The canonical momentum $\Pi_{\phi_n}(t)$ of the field $\phi_{n}(t)$ is 
\beq
\Pi_{\phi_n}(t)=\frac{\delta L}{\delta \left(\frac{d}{dt}\phi_{n} (t)\right)}=i\overline{\phi}_{n}(t),
\eeq
so that the canonical commutation relations read
\beq
[\phi_{n}(t),\phi_{m}(t)]=[\overline{\phi}_{n}(t), \overline{\phi}_{m}(t)]=0, 
\eeq
\beq
[\phi_{n}(t), \overline{\phi}_{m}(t)]=\delta_{n, m}.
\eeq
If the interaction is switched off (setting $\lambda=0$, or moving to the interaction representation), the equation of motion determines the time dependence of the field explicitly, 
\beq
\phi_n(t)~~\overrightarrow{\mbox{\tiny free}}~~e^{-i E_n^b t} a_n,
\eeq
with 
\beq
[a_n, a_m]=[\overline{a}_n, \overline{a}_m]=0,
\eeq
\beq
[a_n, \overline{a}_m]=\delta_{m, n}.
\eeq
From these expressions, $a_n$ and $\overline{a}_n$ are, respectively, understood as the creation and annihilation operators of the particle of the species $n$.

The Hamiltonian of Model $A$ is 
\bea
H_{A} =E_n^b~\overline{\phi}_n(t)\phi_{n}(t) 
	+ \sum_{n_1, n_2, n_3}\lambda \left(\overline{\phi}_{n_1}(t)\phi_{n_2}(t)\phi_{n_3}(t)+\phi_{n_1}(t)\overline{\phi}_{n_2}(t)\overline{\phi}_{n_3}(t)\right).
\eea

Therefore, the free Hamiltonian becomes
\beq
H_{A}^{(0)}=\sum_{n}E_n^b~\overline{\phi}_n(t)\phi_{n}(t)=\sum_{n}E_n^b~\overline{a}_n a_{n}.
\eeq
The energy of the state $|c_1, c_2, c_3, \dots\rangle$, which is obtained from the vacuum $|0\rangle$ by creating $c_1$ particle $n_1$, $c_2$ particle $n_2$, $c_3$ particle $n_3, \dots$, 
\beq
|c_1, c_2, c_3, \dots\rangle^b=\frac{1}{n_1!}(\overline{a}_{n_1})^{c_1}\frac{1}{n_2!}(\overline{a}_{n_2})^{c_2}\frac{1}{n_3!}(\overline{a}_{n_3})^{c_3}\cdots |0\rangle,
\eeq
is given by
\beq
H_{A}^{(0)}|c_1, c_2, c_3, \dots\rangle^b=E_c^b|c_1, c_2, c_3, \dots\rangle^b,
\eeq
where
\beq
E_c^b=c_{1}E_{n_{1}}^b+c_{2}E_{n_{2}}^b+c_{3}E_{n_{3}}^b+\cdots. \label{energy conservation}
\eeq
Here, the vacuum $|0\rangle$ is defined by $a_n |0\rangle=0$ for any $n=1, 2, 3, \dots$, and the affix $``b"$ means ``bosonic" as before. 
This bosonic state corresponds to the factorization of a number $n$ into
\beq
n=(n_1)^{c_1}(n_2)^{c_2}(n_3)^{c_3}\cdots. 
\label{bosonic factorization}
\eeq
If we can restrict the labels of the bosonic fields to be prime numbers, then Eq.(\ref{bosonic factorization}) reproduces precisely Eq.(\ref{factorization into primes}).

Next, we introduce the fermionic field $\psi_{n}(t)$ labeled by a natural number $n$ which interacts with the bosonic field $\phi_{n}(t)$ by Yukawa coupling $y$.  Then, we have ``Model $B$" whose Lagrangian is given by
\bea
	L_{B}&=&\sum_{n}\left(\overline{\phi}_{n}(t)i \frac{d}{dt}\phi_{n}(t)-E_n^b~\overline{\phi}_{n}(t)\phi_{n}(t)\right) \nonumber \\ 
	&+&\sum_{n}\left(\overline{\psi}_{n}(t)i \frac{d}{dt}\psi_{n}(t)-E_n^f~\overline{\psi}_{n}(t)\psi_{n}(t)\right) \nonumber \\ 
	&-&	\sum_{n_1, n_2, n_3}y\left(\overline{\psi}_{n_1}(t)\psi_{n_2}(t)\right) \left(\phi_{n_3}(t)+\overline{\phi}_{n_3}(t)\right) \nonumber \\ 
	&-&\sum_{n_1, n_2, n_3}\lambda \left(\overline{\phi}_{n_1}(t)\phi_{n_2}(t)\phi_{n_3}(t)+\phi_{n_1}(t)\overline{\phi}_{n_2}(t)\overline{\phi}_{n_3}(t)\right).
\eea
Here, the affix ``$f$" means ``fermionic".

The Hamiltonian of Model $B$ is
\bea
	H_{B}&=&\sum_{n}\left(E_n^b\left(\overline{\phi}_{n}(t)\phi_{n}(t)\right)+E_n^f\left(\overline{\psi}_{n}(t)\psi_{n}(t)\right) \right)\nonumber \\ 
	&+&\sum_{n_1, n_2, n_3} y \left(\overline{\psi}_{n_1}(t)\psi_{n_2}(t)\right)\left(\phi_{n_3}(t)+\overline{\phi}_{n_3}(t)\right) \nonumber \\ 
	&+&\sum_{n_1, n_2, n_3}\lambda \left(\overline{\phi}_{n_1}(t)\phi_{n_2}(t)\phi_{n_3}(t)+\phi_{n_1}(t)\overline{\phi}_{n_2}(t)\overline{\phi}_{n_3}(t)\right),
\eea
and its free part reads
\beq
H_{B}^{(0)}=\sum_{n}\left(E_n^b\left(\overline{\phi}_{n}(t)\phi_{n}(t)\right)+E_n^f\left(\overline{\psi}_{n}(t)\psi_{n}(t)\right)\right).
\eeq

The quantization of the fermionic field, $\psi_{n}(t)$, is done similarly as the bosonic field, but anti-commutator is used rather than the commutator in the former case.  Since the conjugate momentum of $\psi_{n}(t)$ is
\beq
\Pi_{\psi_n}(t)=\frac{\delta L}{\delta\left(\frac{d}{dt}\psi_{n} (t)\right)}=i\overline{\psi}_{n}(t),
\eeq
the canonical commutation relations read
\beq
\{\psi_{n}(t),\psi_{m}(t)\}=\{\overline{\psi}_{n}(t), \overline{\psi}_{m}(t)\}=0, 
\eeq
\beq
\{\psi_{n}(t), \overline{\psi}_{m}(t)\}=\delta_{n, m}.
\eeq
If the interaction is switched off (or in the interaction representation), we have
\beq
\psi_n(t)~~\overrightarrow{\mbox{\tiny free}}~~e^{-i E_n^f t} b_n,
\eeq
with 
\beq
\{b_n, b_m\}=\{\overline{b}_n, \overline{b}_m\}=0,
\eeq
\beq
\{b_n, \overline{b}_m\}=\delta_{m, n}.
\eeq

From the anti-commutation relations, creation of fermionic particle can not be doubled, so that the state with the fermionic excitation takes the following form
\beq
|d_1, d_2, d_3, \dots\rangle^f=\overline{b}_{n_1}\overline{b}_{n_2}\overline{b}_{n_3}\cdots |0\rangle.
\eeq
The energy of the state (eigen-value for $H_B^{0}$) is 
\beq
E_d^f=E_{n_{1}}^f+E_{n_{2}}^f+E_{n_{3}}^f+\cdots. \label{fermionic factorizaion}
\eeq
 Therefore, the fermionic state represents the natural number $n$ decomposed into
\beq
n=n_1n_2n_3\cdots. \label{fermionic factorization}
\eeq

If we restrict the labels of the ferimonic fields to be prime numbers, then Eq.(\ref{fermionic factorizaion}) reproduces the number $n$, which is not divisible by any square of prime numbers other than 1.  For this kind of number $n$, M\"obius function $\mu(n)$ is introduced in number theory in order to count the number of prime factors in $n$.  This function is called fermion number $F$ in physics, representing the number of fermions excited from the vacuum, where each fermion can not be excited repeatedly.  Then, $(-1)^{\mu}$ is the fermionic parity $(-1)^F$ in physics.  Number theory seems to be coming and going between the bosonic representaion and the fermionic representation (see p.8 of \cite{Riemann}), so that the ``supersymmetry (SUSY)", the symmetry exchanging bosons and fermions ($\phi(t)\leftrightarrow\psi(t)$) and is the target of the next collider experiments such as LHC and ILC, seems to play an important role also in number theory. 

In order to have the supersymmetric (SUSY) model ``Model $C$", we may tune the parameters in Model $B$ as follows:
\bea
E_n^b=E_n^f~~(\rm {or}~~\omega^b=\omega^f=\omega), \rm{and}~~~\lambda=y. \label{tuning of parameters}
\eea

In quantum mechanics or in quantum field theory, one state changes to another state (making transition) in the lapse of time. The conservation of energy is not guaranteed at finite lapse of time because of uncertainty principle in quantum theory, but it holds at infinite lapse of time.  Therefore, we have to consider the transition at infinite lapse of time also in number theory to gurantee the energy conservation.  The transition probabitlity from an initial sate $|i\rangle$ to an final state $|f\rangle$ at infinite lapse of time is called S-matrix element and is given by
\beq
\langle f |S| i \rangle=\delta_{fi}+i \langle f |T| i \rangle,
\eeq
\bea
&&\langle f |T| i \rangle=\langle f|H^{(I)}|i\rangle+
\sum_{m(\ne i)} \langle f|H^{(I)}|m\rangle \frac{1}{E_i-E_m+i\epsilon}\langle m|H^{(I)}|i\rangle \nonumber \\ 
&+&\sum_{m_1, m_2(\ne i)} \langle f|H^{(I)}|m_1\rangle \frac{1}{E_i^{(0)}-E_{m_1}^{(0)}+i\epsilon}\langle m_1|H^{(I)}|m_2\rangle \frac{1}{E_i^{(0)}-E_{m_2}^{(0)}+i\epsilon}\langle m_2|H^{(I)}|i\rangle \nonumber \\ 
&+&\cdots, \label{old fashioned pert}
\eea
where $H^{(I)}$ is the interaction Hamiltonian, the remaining part of the Hamiltonian other than the free Hamiltonian $H^{(0)}$, which depends on the coupling constants $\lambda$ and $y$, while $E_a^{(0)}$ is the energy eigen-value of the state $|a\rangle$ for the free Hamiltonian $H^{(0)}$.

The energy eigen-value (energy level) is usually shifted by the interactions,
\bea
E_n&=&E_n^{(0)}+\langle n|H^{(I)}|n\rangle +\langle n|H^{(I)}|m\rangle\frac{1}{E_n^{(0)}-E_m^{(0)}+i\epsilon}\langle m|H^{(I)}|n\rangle \nonumber \\ 
&+&\cdots.
\eea

However, in order to study how a natural number factorizes into other (prime) numbers, the shift of energy level from the original value $E_n^{(0)}=\omega \ln n$ is unfavorable.  In this respect the supersymmetric model such as Model $C$ is promissing, since in the supersymmetric model the shift of energy level does not usually occur owing to the cancellation between bosonic and fermionic intermediate states.

Now, we will discuss the decay of particle.  Since S-matrix is unitary
\beq
\sum_m \langle f|S^{\dagger}|m\rangle \langle m|S|i\rangle=\delta_{fi},
\eeq
the so-called optical theorem holds, 
\beq
\Gamma_{\rm{total}}(n)=\sum_m \vert \langle m|T|n \rangle \vert^2=2 Im\langle n|T|n \rangle,
\eeq
where the ``total decay width" $\Gamma_{\rm{total}}(n)$ means the probablity of the particle $n$ decaying into all kinds of final sets of particles, and is itentical to the imaginary part of the forward ({\it i.e.} initial and final state are  identical) scattering amplitude.

Therefore, whether a natural number $n$ is prime number or not can be determined by the analytic property of the forward scattering amplitude as follows:
\bea
Im\langle n|T|n \rangle&=&0,~~\mbox{if  $n$ is prime},\\
&\ne&0,~~\mbox{if n is not prime}.
\eea

Hence the number of prime numbers that are smaller than a given quantity $x$, that is $F(x)$ in \cite{Riemann}, can be given as
\bea
F(x)=\sum_{n=1}^{[x]}\delta_{\{Im\langle n|T|n \rangle, 0\}},
\eea
where $[x]$ is the Gauss symbol, giving the maximum integer less than $x$.  The expression may be approximated for large $x$ as
\bea
F(x) \approx \int_1^x dy~\delta(Im\langle y|T|y \rangle).
\eea
Therefore, if the proper expression of the decay width or the forward scattering amplitudes is obtained, perturbatively or non-perturbatively, the distribution function of prime numbers, $F(x)$, can be estimated.

\section{String-like model giving logarithmic series of energies}
In this section we will propose string-like models, giving automatically the logarithmic series of energy, $E_n=\omega\ln n$.

The simplest example is ``Model $A'$", which reproduces Model $A$ of the last section. Its Lagrangian density is
\bea
{\cal L}_{A'}&=&\overline{\phi}(t, \sigma) i \frac{\partial}{\partial t}\phi(t, \sigma)-\overline{\phi}(t, \sigma)\omega\ln \left(1-i\frac{\partial}{\partial \sigma}\right)\phi(t, \sigma) \nonumber \\ 
	&-&(2\pi)^{3/2}\lambda \delta(\sigma)\left(\overline{\phi}(t, \sigma)\phi(t, \sigma)\phi(t, \sigma)+\phi(t, \sigma)\overline{\phi}(t, \sigma)\overline{\phi}(t, \sigma)\right),
\eea
where the bosonic field $\phi(t, \sigma)$ is assumed to satisfy the periodic boundary condition, \beq
\phi(t, \sigma)=\phi(t, \sigma+2\pi).
\eeq
So the physical object described by this model is a string-like one of the closed type, being parametrized by $\sigma$.

Using the Fourier expansion of  $\phi(t, \sigma)$ 
\beq
\phi(t, \sigma)=\sum_{n=0}^{\infty}\frac{1}{\sqrt{2\pi}}\phi_{n+1} (t) e^{+in\sigma},
\eeq
the Lagrangian of the Model $A'$ reproduces the Model $A$, that is,
\beq
\int_0^{2\pi} d\sigma {\cal L}_{A'}(t, \sigma)=L_A(t).
\eeq

Therefore, Model $A'$ is the field theoretical realization of the Model $A$.

Here, we have to comment on what kind of set of fields has to be prepared.  In the Model $A'$, the set of fields prepared is labeled by the whole integers, $n=1, 2, 3, 4, 5, 6, \dots$.  It is of course better to prepare the set of fields labeled by the whole prime numbers, $p=1, 2, 3, 5, 7, \dots$ for the study of number theory.  To prepare the latter set in physics, however, we have to consider the more elaborate dynamical system, such as Chaos {\it e.t.c.}.  On the other hand, the set of fields labeled by the natural integers is easier to prepare, but even with this set of fields prime or not prime can be understood as the elementary nature of particle, or as decay or not decay of particles, so that we will adopt this simple set of fields, labeled by the natural integers in this paper.

Next, we will reproduce the Model $B$ by a string-like ``Model $B'$".  Its Lagrangian density is given by
\bea
	{\cal L}_{B'}&=&\overline{\phi}(t, \sigma)i \frac{\partial}{\partial t}\phi(t, \sigma)-\overline{\phi}(t, \sigma)\omega^b \ln \left(1-i\frac{\partial}{\partial \sigma}\right)\phi(t, \sigma) \nonumber \\ 
	&+&\overline{\psi}(t, \sigma)i \frac{\partial}{\partial t}\psi(t, \sigma)-\overline{\psi}(t, \sigma)\omega^f \ln \left(1-i\frac{\partial}{\partial \sigma}\right)\psi(t, \sigma) \nonumber \\ 
	&-&(2\pi)^{3/2} y \delta(\sigma)\left(\overline{\psi}(t, \sigma)\psi(t, \sigma)\right) \left(\phi(t, \sigma)+\overline{\phi}(t, \sigma)\right) \nonumber \\ 
	&-&(2\pi)^{3/2}\lambda \delta(\sigma)\left(\overline{\phi}(t, \sigma)\phi(t, \sigma)\phi(t, \sigma)+\phi(t, \sigma)\overline{\phi}(t, \sigma)\overline{\phi}(t, \sigma)\right), 
\eea
where the fermionic field $\psi(t, \sigma)$ is also assumed to be periodic\footnote{Anti-periodic  boundary condition is also possible for the fermionic field.},
\beq
\psi(t, \sigma)=\psi(t, \sigma+2\pi).
\eeq

Similarly, Fourier expansion of  $\phi(t, \sigma)$ 
\beq
\psi(t, \sigma)=\sum_{n=0}^{\infty}\frac{1}{\sqrt{2\pi}}\psi_{n+1}(t) e^{+in\sigma},
\eeq
reproduces the Lagrangian of the Model $B$ as
\beq
\int_0^{2\pi} d\sigma {\cal L}_{B'}(t, \sigma)=L_B(t).
\eeq

The supersymmetric (SUSY) string-like ``Model $C'$" can be also introduced by tuning the parameters in Model $B'$ as in Eq.(\ref{tuning of parameters}).

\section{One-loop vacuum amplitude, free energy and zeta function}
Scattering amplitudes can be estimated using the old-fashioned perturbation expressed in Eq.(\ref{old fashioned pert}), but are obtained more easily by the path integration method of Feynman.
For example, the partition function $Z_A[J]$ and the connected generating functional $W_A[J]$ at zero temperature ($T=0$) of Model $A'$ are given by introducing the external source $J$ as
\bea
Z_A[J]=e^{i W_A[J]}=\int {\cal D}\phi(\xi){\cal D}\overline{\phi}(\xi) \exp i \int dt d\sigma \left({\cal L}_{A'}+\overline{J}(\xi)\phi(\xi)+J(\xi)\overline{\phi}(\xi)\right),
\eea
Here, $(\xi)$ denotes $(t, \sigma)$.
Successive functional differentiation by sources determines the connected N-point function (amplitude) for $\phi(\xi)$,
\bea
\langle \phi(\xi_1), \phi(\xi_2), \dots, \phi(\xi_N)\rangle=\frac{\delta^N W_A[J]}{i^N \delta J(\xi_1)\delta J(\xi_2)\cdots \delta J(\xi_N)} \Bigg{\vert}_{J=0}
\eea  
This N-point function is directly connected to the  N-point scattering amplitude by the so-called reduction formula of Lehmann, Symanzik, and Zimmermann (LSZ).

First, we examine the one-loop connected vacuum ($N=0$) amplitude at $T=0$, since for this case we need not consider the interactions as well as the external sources.
\bea
iW_A(\mbox{1-loop vac})_{T=0}
=\int_0^{\infty} \frac{db}{b} \ln \int_{\phi(t, \sigma)=\phi(t+b,\sigma)} {\cal D}\phi(\xi){\cal D}\overline{\phi}(\xi) \exp i \int dt d\sigma {\cal L}_{A'}.
\eea
Here, one-loop means the bosonic (fermionic) field propagates and comes back (with opposite sign) to its original value after the lapse of time $b$, forming a loop, and so we impose the condition,
\beq
\phi(t, \sigma)=\phi(t+b,\sigma),
\eeq
in addition to the periodic boundary condition for the closed string-like object,
\beq
\phi(t, \sigma)=\phi(t,\sigma+2\pi).
\eeq
Then, the parameter space $(t, \sigma)$ becomes a torus, having the modular symmetry of $SL(2, Z)$.

Using the mode expansion of the field
\beq
\phi(t, \sigma)=\sum_{m=-\infty}^{\infty}\sum_{n=0}^{\infty}\frac{1}{\sqrt{2\pi b}} a_{m, n+1}e^{-i\left(\frac{2\pi}{b} mt-n\sigma\right)},
\eeq
and performing the Gaussian integration, we obtain\beq
iW_A(\mbox{1-loop vac})_{T=0}
=\int_0^{\infty} \frac{db}{b} \sum_{m=-\infty}^{\infty}\sum_{n=0}^{\infty}-\ln \Bigg{|}\left(\frac{2\pi}{b} m -\omega\ln (1+n)\right)\Bigg{|},
\eeq
where $\ln (\det A)^{-1}=-Tr \ln A$ has been used.
Summation over $m$ is carried out by
\beq
\sum_{m=-\infty}^{\infty}\ln (m+X)=\int_0^X dy \sum_{m=-\infty}^{\infty}\frac{1}{m+y}=\ln(\sin \pi X),
\eeq
where the irrelevant constants are ignored, then we have 
\bea
iW_A(\mbox{1-loop vac})_{T=0}
=-\sum_{n=1}^{\infty}\int_0^{\infty} \frac{db}{b} \ln \left(\sin \left(\frac{b\omega}{2}\ln n\right)\right).
\eea

The free energy (Helmholtz free energy) at finite temperature is almost similar quantity as the one-loop vacuum amplitude at zero temperature.  At finite temperature vacuum is replaced by the heat bath.  Usually this replacement is done by making time $t$ to be imaginary ($t=i\tau$), and  
the periodicity is imposed on the interval $0<\tau<\beta$, in the similar manner as the one-loop case.  Here, we have to fix $\beta$, corresponding to fixing the temperature, and do not integrate over $\beta$.
This $\beta$ can be identified to the inverse temperature $T^{-1}$ (more precisely, $\hbar/\beta=k_B T$, when recovering Planck constant/$2\pi=\hbar$ and Boltzmann constant=$k_B$).  The validity of this replacement can be understood from that the quantum mechanical time evolution operator becomes the Boltzmann weight in the finite temperature statistical mechanics, namely,
\bea
e^{i \int dt H }\rightarrow e^{(i )^2 \int_0^{\beta} d\tau H}=e^{-\beta H}=e^{-\frac{1}{T}H}
\eea

Then, the free energy $F_A$ of the Model $A'$ is given by the partition function as $F_A=-\frac{1}{\beta}\ln Z_A$.

Now, we can obtain the free energies at finite temperature of our Models $A^{(\prime)}$, $B^{(\prime)}$, and $C^{(\prime)}$, in the case without interactions, as follows:
 \bea
F_A(\beta)
&=&\frac{1}{\beta}\sum_{n=1}^{\infty} \ln \left(\sinh \left(\frac{\beta\omega^{b}}{2}\ln n\right)\right). \nonumber \\
&=&\sum_{n=1}^{\infty}  \left\{ \frac{1}{2}\omega^{b} \ln n + \frac{1}{\beta} \ln \left(1- e^{-\beta \omega^{b} \ln n} \right) \right\}, \\
F_B(\beta)
&=&F_A(\beta)-\frac{1}{\beta}\sum_{n=1}^{\infty} \ln \left(\cosh \left(\frac{\beta\omega^{f}}{2}\ln n\right)\right). \nonumber \\
&=&\sum_{n=1}^{\infty}  \left\{ \frac{1}{2}(\omega^{b}- \omega^{f})\ln n + \frac{1}{\beta} \ln \left(\frac{1- e^{-\beta \omega^{b} \ln n}}{1+ e^{-\beta \omega^{f} \ln n}}\right) \right\},\\
F_{C}(\beta)
&=&\frac{1}{\beta} \sum_{n=1}^{\infty} \ln \left( \frac{1- e^{-\beta \omega \ln n}}{1+ e^{-\beta \omega \ln n}} \right).
\eea

The corresponding partition functions of Models $A'$, $B'$, and $C'$ are
\bea
Z_A(\beta)&=&\prod_{n=\mbox{\tiny positive integers}} n^{-\frac{1}{2}\beta \omega^{b}} \frac{1}{1-n^{-\beta \omega^{b}}}, \\
Z_B(\beta)&=&\prod_{n=\mbox{\tiny positive integers}} n^{-\frac{1}{2}\beta (\omega^{b}-\omega^{f})}
\frac{1+n^{-\beta \omega^{f}}}{1-n^{-\beta \omega^{b}}}, \\
Z_C(\beta)&=&\prod_{n=\mbox{\tiny positive integers}} \frac{1+n^{-\beta \omega}}{1-n^{-\beta \omega}}.
\eea

The zeta function is a kind of partition function, being expressed as
\bea
\zeta(s)=\sum_{n=\mbox{\tiny positive integers}}n^{-s}=\prod_{p=\mbox{\tiny prime numbers}}\left(\frac{1}{1-p^{-s}}\right),
\eea
so that the parameter $s$ be $\beta\omega=\omega/T$ in physics, namely the energy unit of the system $\hbar \omega$ measured in the energy unit of the heat bath $k_B T$.  Therefore, the discrete symmetry between $s \leftrightarrow (1-s)$, studied by Riemann \cite{Riemann} means the symmetry between the hot heat bath ($k_B T > \hbar \omega$) and the cold heat bath ($k_B T < \hbar \omega$).  The symmetry is established by using Jacobi's identity, or in physics, by using the dual transformation from the model defined on an integer lattice to the model defined on its dual lattice, the physical meaning of this symmetry is expected to be clarified.  The dynamical system behind the zeta function is similar to Model $A$, but there is a difference.  Contrary to Model $A$, the zeta function has no zero point energy (without having the factor $n^{-\frac{1}{2}s}$) and the set of fields are labeled, not by integers but by prime numbers.  The reason why $-\frac{1}{s} \ln \zeta (s)$ plays an important role in \cite{Riemann} is that it is the free energy of the dynamical system behind the zeta.

The zeta-like function relevant to Model $B$ is
\bea
\zeta^{f}(s)=\sum_{n=\mbox{\tiny fermionic numbers}}n^{-s}=\prod_{p=\mbox{\tiny prime numbers}}(1+p^{-s}),
\eea
where the affix ``f'' means ``fermionic'' and the ``fermionic number'' means the number not divisible by the square of prime numbers.  The corresponding zeta function for Model C is $\zeta(s)\zeta^{f}(s)$.

The partition function $Z(\beta)$ and the free energy $F(\beta)$ are in general 
\bea
Z(\beta)=Tr~e^{-{\hat H}}, 
F(\beta)=-\frac{1}{\beta} \ln Z(\beta),
\eea
where ${\hat H}$ is the Hamiltonian operator of the system.
If we replace the trace by the supersymmetric trace $Str$, then we have
\bea
{\tilde Z}(\beta)=Str~e^{-{\hat H}}=Tr~(-1)^F e^{-{\hat H}}, 
{\tilde F}(\beta)=-\frac{1}{\beta} \ln {\tilde Z}(\beta)
\eea

The example of this is
\bea
\frac{1}{\zeta(s)}= \sum_{\mbox{\tiny positive integers}} (-1)^{\mu} n^{-s},
\eea
and for the supersymmetric (SUSY) model $C$, ${\tilde Z}_C(\beta)=1$ and ${\tilde F}_C(\beta)=0$ becomes trivial.  This means the partition function and the free energy defined by $Str$ is a good measure to observe SUSY breaking.  This $Str$ appears also in the one-loop amplitude at zero temperature, since each particle contributes to the loop amplitude with the fermionic parity $(-1)^F$.  Now, the one-loop vacuum amplitude for the Model $B$ and $C$ reads
\bea
iW_B(\mbox{1-loop vac})_{T=0}
&=&-\sum_{n=1}^{\infty}\int_0^{\infty} \frac{db}{b}  \ln \left(\frac{\sin \left(\frac{b\omega^b}{2}\ln n\right)}{\sin \left(\frac{b\omega^f}{2}\ln n\right)}\right), \\
iW_{C}(\mbox{1-loop vac})_{T=0}
&=&0,
\eea
where it it noted that $(-1)^F$ alters ``cos" to ``sin". 
Now, the vanishing of the one-loop vacuum amplitude in Model $C'$, $W_{C}=0$, can be shown, which is, however, an entrance to the general statement in SUSY model that there is no radiative corrections to the quadratic mass terms of the fields.  This general statement is expected to hold also in our Model $C$ with the proof.

\section{Symmetry of the dynamical system} 
One of the merit of introducing the string-like models (Model $A'$, Model $B'$, Model $C'$, {\it etc.}) is to elucidate the symmetry of the dynamical system, such as $SL(2, {\bf Z})$, which is implicit and hidden in the quantum mechanical models (Model $A$, Model $B$, Model $C$, {\it etc.}), even though the latters are equivalent to the formers.

As was stated in the last section, the symmetry of parameter (coordinate) space $(\tau, \sigma)$ which appears in the one-loop vacuum amplitues or the free energies of the string-like models is $SL(2, {\bf Z})$, the symmetry of the torus,
\bea
\left( \begin{array}{c}
t'/b  \\
\sigma'/2\pi
\end{array}\right)=
\left( \begin{array}{cc}
\alpha&\beta  \\
\gamma&\delta
\end{array}\right)
\left( \begin{array}{c}
t/b  \\
\sigma/2\pi
\end{array}\right), \label{coordinate transformation}
\eea
where the matrix $T$ 
\bea
T=\left( \begin{array}{cc}
\alpha&\beta  \\
\gamma&\delta
\end{array}\right)
\eea
is an element of $SL(2, {\bf Z})$.  Corresoponding to the coordinate transformation, we have 
\bea
\left( \begin{array}{c}
b \partial_{t'} \\
2\pi \partial_{\sigma'}
\end{array}\right)=
T^{-1}
\left( \begin{array}{c}
b \partial_t  \\
2\pi \partial_{\sigma}
\end{array}\right),  \label{transformation of derivatives}
\eea
where $(b, t)$ are understood to be $(i\beta, i\tau)$ for the free energies.

If number theory is considered as a dynamical system in physics, its symmetry appears in the action or in the free energy.  Under $SL(2, {\bf Z})$ transformation of coordinates, the fields are assumed to transform as scalars,
\bea
\phi(\xi) \rightarrow \phi'(\xi')=\phi(\xi),~~\mbox{and}~~
\psi(\xi) \rightarrow \psi'(\xi')=\psi(\xi).
\eea
This is the ordinary treatment of the coordinate transformation such as the general coordinate transformation (the differomorphism) in physics. 
Then, the free energy of the Molel $A'$, for example, transforms as
\bea
F_A(\beta) \rightarrow F_A(\beta)'=\frac{1}{\beta}\sum_{m,n}\ln \left(\frac{2\pi}{b}m'-\omega^{b}\ln(1+n') \right), \label{transformation of free energy}
\eea
where $(m', -n')$ are given by
\bea
\left( \begin{array}{c}
m' \\
-n'
\end{array}\right)=
T^{-1}
\left( \begin{array}{c}
m \\
-n
\end{array}\right).
\eea
For the $SL(2, {\bf Z})$ transformation, if $\sum_{m,n}$ can be replaced by $\sum_{m',n'}$, then the invariance of the free energies under $SL(2, {\bf Z})$ hold.  This is the case when the interactions are switched off (or $\lambda=0$ and $y=0$).

In order for the replacement of $\sum_{m,n}$ by $\sum_{m',n'}$ works correctly, the range of $m$ and $n$ should be both $-\infty<m, n<+\infty$.  So far the range of $n$ is $0<n<+\infty$, the modification of the model is necessary.  This can be done naively by using $\ln (1+n)=\ln(1+|n|)$, but the more careful examination is needed based on the analytic continuation of the logarithm.

When the interactions are turned on, however, the symmetry is broken down to the additive group of integers ${\bf Z}_{+}$ which is defined by imposing $\gamma=0$ and $|\delta|=1$.  The reason is as follows: The interaction terms have $\delta(\sigma)$ which is not invariant under $SL(2, {\bf Z})$, but becomes invariant,
\bea
\delta(\sigma')&=&\delta(2\pi \gamma  t/b+\delta \sigma)=\frac{1}{|\delta|}\delta(\sigma)=\delta(\sigma). 
\eea
when $\gamma=0$ and $|\delta|=1$.  This shows the dynamical system is invariant under ${\bf Z}_{+}$ even after the interactions are turned on.

In order to understand $SL(2, {\bf Z})$ in the same manner as the discrete symmetries such as the parity (P), the charge conjugation (C), and the time reversal (T) , the following treatment is convenient.  This is to transform only the fields, without changing the coordinates.  Namely, $SL(2, {\bf Z})$ transformation is carried out by
\bea
\phi(\tau, \sigma) &\rightarrow& \phi(\tau', \sigma') ,\\
\psi(\tau, \sigma) &\rightarrow& \psi(\tau', \sigma'), \label{field transformation}
\eea
where the $(\tau', \sigma')$ and $(\tau, \sigma)$ are related by Eq.(\ref{coordinate transformation}).  This treatment is, of course, equivalent to the former.

Then, the tranformation can be understood as the transformation of $m$ and $n$,
\bea
SL(2, {\bf Z}): (m, -n) \rightarrow (m', -n')=(m, -n)T. 
\eea

Now, the discrete symmetries of P, C, and T are
\bea
\mbox{P}: (\phi~\mbox{and}~\psi)(\tau, \sigma) \rightarrow (\phi~\mbox{and}~\psi)(\tau', \sigma')&=&(\phi ~\mbox{and}~\psi)(\tau, -\sigma), \\
\mbox{C}: (\phi ~\mbox{and} ~\psi)(\tau, \sigma) \rightarrow (\phi ~\mbox{and} ~\psi)(\tau', \sigma')&=&({\overline \phi} ~\mbox{and}~{\overline \psi})(\tau, \sigma), \\
\mbox{T}: (\phi ~\mbox{and} ~\psi)(\tau, \sigma) \rightarrow (\phi ~\mbox{and} ~\psi)(\tau', \sigma')&=&(\phi ~\mbox{and} ~\psi)(-\tau, \sigma),
\eea
which can be understood as
\bea
\mbox{P}: (m, n) \rightarrow (m', n')&=&(m, -n), \\
\mbox{C}: (m, n) \rightarrow (m', n')&=&(-m, -n), \\
\mbox{T}: (m, n) \rightarrow (m', n')&=&(-m, n).
\eea
Then, the extension of the range of $n$ to $-\infty <n<+\infty$ makes the dynamical system to be invariant also under $P$ and $C$, so that the concept of the ``anti-paritcle" labeled by $-n$, the charge conjugate state of the particle labeled by $n$, appears here.  The energy levels of particle and anti-particle are usually identical.  This implies the above mentioned naive $\ln (1+n)=\ln(1+|n|)$, but the analytic continuation often requires the additional $\pm i\pi$, making the imaginary part of the energy to be asymmetric, or introducing the unusual structure in the Fermi (as well as the Bose) surface.  

\section{Conclusion}
In this paper, factorization of a number into prime numbers is considered as a particle decays into elementary particles conserving energy.  For this purpose, we have to prepare infinite spieces of particles having energy $E_n=\omega \ln n~(n=1, 2, 3, 4, ...)$.  These infinite spieces of particles are combined to a single field, giving the stirng-like models. By introducing the proper interactions, we can understand whether a number is prime or not by the fact that the corresponding particle can or can not decay. This is a story at zero temperature.  

After moving to finite temperature, and estimating the partition functions and the (Helmholtz) free energies, we can understand the (Riemann) zeta function is a kind of partition function for the dynamical model consisting of a set of fields labeled by prime numbers. The models proposed in this paper, however, consist of a set of fields labeled, not by prime numbers but by integers.

Models made of fields labeled by prime numbers are more direct in the context of primeness.  On the other hand the models proposed in this paper are made of fields labeled by integers, so that they are less direct in this primeness. However, the models are field theoretically constructed and the interactions are introduced naturally to study the transition (or decay) probablity of particles, so that the deficit above may be covered.  For example, the distribution function of prime numbers can be estimated, if the decay widths or the forward scattering amplitudes of particles are known.

The zeta function is a kind of finite temperature partition function of the dynamical system.  We know the finite temperature statistical mechanics is a powerful method to study the averaged or the smeared property of the system surrounded by the heat bath.  By the existence of the heat bath, the energy conservation is violated, since the heat bath emits or absorbs energy.  Therefore, in additon to this averaged or smeared study of number theory using the free energies or the zeta functions, we wish to utilize the zero temperature field theory in which  the energy is always conserved and the primeness is also recognized by the transition probability.  Therefore, going back and forth between zero and non-zero temperature studies is necessary. In these studies, the supersymmetric (SUSY) models such as Model $C'$, may be interesting, since the supersymmetric model has the merit on the stability of energy levels against radiative corrections and the cancellation of zero point energies.


%
%
\section*{Acknowledgements}
The author gives his sincere thanks to his family for their understanding and supporting of this work.

%
%

\end{document}